\documentclass[sigconf,nonacm]{style/acmart}

\AtBeginDocument{%
  }

\usepackage{amsmath}
\usepackage{multirow}
\usepackage{tabularx}
\usepackage{pifont}
\usepackage{wasysym}
\DeclareFontFamily{U}{wasy}{}
\DeclareFontShape{U}{wasy}{m}{n}{<->wasy10}{}
\DeclareFontShape{U}{wasy}{b}{n}{<->ssub*wasy/m/n}{}

\AtBeginDocument{\let\balance\relax}
\usepackage[inline]{enumitem}
\usepackage{xspace}
\usepackage{tikz}
\usetikzlibrary{shapes,arrows}
\usetikzlibrary{backgrounds, positioning, fit}
\usepackage{fvextra}
\input{pygmented/pygstyle.tex}
\usepackage[most]{tcolorbox}

\graphicspath{{figures/}}

\definecolor{inkslate}{HTML}{1E2A40}
\definecolor{okgreen}{HTML}{005C30}
\definecolor{errred}{HTML}{A72D26}
\definecolor{emphviolet}{HTML}{6F00A2}
\definecolor{washneutral}{HTML}{D5D8DE}
\definecolor{washinfo}{HTML}{DDEBFF}
\definecolor{washseed}{HTML}{F8E9D1}

\newtcolorbox{callout}[2][]{colback=#2, colframe=inkslate!30, boxrule=0.4pt,
  arc=1.5pt, left=5pt, right=5pt, top=4pt, bottom=4pt, fontupper=\normalsize, #1}
\newenvironment{takeaway}{\begin{callout}{washneutral}\noindent\textbf{Takeaway.}\enspace\ignorespaces}{\end{callout}}

\newcommand{\method}{\textsc{Amulet}\xspace}

\newcommand{\race}{\texttt{Race}\xspace}
\newcommand{\sex}{\texttt{Sex}\xspace}

\newcommand{\dtrain}{\mathcal{D}_{tr}\xspace}

\newcommand{\model}{\mathcal{M}\xspace}
\newcommand{\modelstd}{\mathcal{M}_{std}\xspace}
\newcommand{\modeldef}{\mathcal{M}_{def}\xspace}
\newcommand{\modelstol}{\mathcal{M}_{stol}\xspace}
\newcommand{\modelpois}{\mathcal{M}_{pois}\xspace}

\newcommand{\fmnist}{\texttt{FMNIST}\xspace}
\newcommand{\lfw}{\texttt{LFW}\xspace}
\newcommand{\cifar}{\texttt{CIFAR10}\xspace}
\newcommand{\census}{\texttt{CENSUS}\xspace}
\newcommand{\celeba}{\texttt{CELEBA}\xspace}
\newcommand{\sstwo}{\texttt{SST-2}\xspace} 

\newcommand{\llama}{\texttt{Llama-3.2-3B}\xspace} 

\newcommand{\smark}{$\LEFTcircle$\xspace}
\newcommand{\xmark}{$\Circle$\xspace}
\newcommand{\cmark}{$\CIRCLE$\xspace}

\begin{document}

\title[\method]{\method: A Python Library for Assessing Interactions Among ML Defenses and Risks}

\author{Asim Waheed}
\authornote{Corresponding author.}
\affiliation{%
  \institution{University of Waterloo}
  \city{Waterloo}
  \state{Ontario}
  \country{Canada}}
\email{a7waheed@uwaterloo.ca}

\author{Vasisht Duddu}
\affiliation{%
  \institution{University of Waterloo}
  \city{Waterloo}
  \state{Ontario}
  \country{Canada}}
\email{vasisht.duddu@uwaterloo.ca}

\author{Sebastian Szyller}
\authornote{Work done while at Intel Labs.}
\affiliation{%
  \institution{Aalto University}
  \city{Espoo}
  \country{Finland}}
\email{contact@sebszyller.com}

\begin{abstract}
Machine learning (ML) models are susceptible to various risks to security, privacy, and fairness.  
Most defenses are designed to protect against each risk individually (or \emph{intended interactions}), but can inadvertently affect susceptibility to other unrelated risks (i.e., \emph{unintended interactions}).  
We introduce \method, the first Python library for evaluating both intended and unintended interactions among ML defenses and risks.  
\method is: 
\begin{enumerate*}[label=\roman*),itemjoin={\xspace}] 
\item \emph{comprehensive} by including representative attacks, defenses and metrics,
\item \emph{extensible} to new modules due to its modular design,
\item \emph{consistent} with a user-friendly API template for inputs and outputs, and
\item \emph{applicable} to evaluate novel interactions.
\end{enumerate*} 
\method spans models from small vision networks such as VGG to billion-parameter large language models.
By satisfying all four requirements, \method offers a unified foundation for the first systematic evaluation of unintended interactions across multiple risks.
\end{abstract}



\maketitle

\section{Introduction}\label{sec:introduction}

Machine learning (ML) models are used in high-stakes decision-making, such as healthcare, hiring, and loan approval.  
These deployments require key properties: \emph{security}, \emph{privacy}, and \emph{fairness}~\cite{duddu2025CombiningMachineLearning, szyller2023ConflictingInteractionsProtection, liu2022MLDoctorHolisticRisk, nicolae2018AdversarialRobustnessToolbox, weerts2023FairlearnAssessingImproving}. 
Also, regulators in several jurisdictions are drafting ML frameworks that require assessing model susceptibility to different risks~\cite{whitehouse2020GuidanceRegulationArtificial, kazim2021AIAuditingImpact, europeanunionlaw2018GeneralDataProtection}.
However, ML models face \emph{risks} that violate these properties~\cite{duddu2024SoKUnintendedInteractions, gittens2022AdversarialPerspectiveAccuracy, chen2023PrivacyFairnessFederated}.  
Various \emph{attacks} have been proposed to realize each of these risks, while \emph{defenses} have been proposed to protect against them~\cite{abadi2016DeepLearningDifferential,maini2021DatasetInferenceOwnership,li2021AntiBackdoorLearningTraining, dwork2012FairnessAwareness}.  
We call the interaction between a defense and the risk it was designed to protect against an \emph{intended interaction}.  
However, a defense for one risk may inadvertently raise or lower susceptibility to other unrelated risks, an \emph{unintended interaction}~\cite{duddu2024SoKUnintendedInteractions}. 

Prior works have proposed a systematic approach to identify the causes underlying interactions among defenses and risks~\cite{duddu2024SoKUnintendedInteractions}. 
However, we need an easier way to evaluate susceptibility to both intended and unintended interactions at scale.   

We identify four key requirements for such a library: it should be \emph{comprehensive}, \emph{extensible}, \emph{consistent}, and \emph{applicable}, which we describe in more detail in Section~\ref{sec:problem}.  
Existing libraries satisfy some, but not all, of these requirements.  
They either focus on a single risk (e.g.~\cite{wei2023DPMLBenchHolisticEvaluation,yousefpour2021OpacusUserFriendlyDifferential,kokhlikyan2020CaptumUnifiedGeneric}), omit unintended interactions (e.g.~\cite{papernot2016TechnicalReportCleverHans,nicolae2018AdversarialRobustnessToolbox,ling2019DEEPSECUniformPlatform,liu2022MLDoctorHolisticRisk,murakonda2020MLPrivacyMeter,bellamy2018AIFairness360,weerts2023FairlearnAssessingImproving}), or are difficult to extend with newer attacks or defenses (e.g.~\cite{ling2019DEEPSECUniformPlatform,pang2022TROJANZOOUnifiedHolistic}).  

We propose \method, a library that meets all the requirements.  
It includes attacks, defenses, and metrics across all three properties: \emph{security} (evasion, poisoning, unauthorized model ownership), \emph{privacy} (inference attacks), and \emph{fairness} (bias).  
We envision \method being used by:
\begin{enumerate*}[label=\roman*),itemjoin={,\xspace}] 
    \item \emph{researchers} to explore new unintended interactions, study variants of attacks, defenses, and metrics, and design effective defenses, and
    \item \emph{practitioners} to analyze trade-offs when deploying ML defenses and inform training decisions.
\end{enumerate*}
Our main contributions are as follows: we
\begin{enumerate}[leftmargin=*,wide,labelindent=0pt]
\item present \method\footnote{\url{https://github.com/ssg-research/amulet}}, the \emph{first} library to evaluate \emph{both} intended and unintended interactions, at scale.
\item characterize its design, showing \method is \emph{comprehensive} (spans diverse risks) and \emph{consistent} (provides uniform API templates) (Section~\ref{sec:design}).
\item demonstrate its \emph{extensibility} by adding a text modality and evaluating interactions on a large language model (LLM) victim (Section~\ref{sec:extensibility}).
\item demonstrate its \emph{applicability} by evaluating unexplored interactions (Section~\ref{sec:evaluation}).

\end{enumerate}

\section{Background and Related Work}\label{sec:background}

We introduce different risks to machine learning (ML) models (Section~\ref{sec:backrisk}), the corresponding defenses (Section~\ref{sec:backdef}), and the interactions between them (Section~\ref{sec:interactions}).

\subsection{Risks against Machine Learning}\label{sec:backrisk}

Following the systematization in prior work~\cite{duddu2024SoKUnintendedInteractions},
we present
risks to security (\ref{evasion},~\ref{poison}, and~\ref{modelext}),
risks to privacy (\ref{meminf},~\ref{attinf},~\ref{datarecon}, and~\ref{distinf}),
and discriminatory behavior (\ref{disc}).
\begin{enumerate}[label=\textbf{S\arabic*},leftmargin=*,wide, labelindent=0pt, itemsep=0pt, topsep=0pt]
\item\label{evasion} \textbf{Evasion} attacks attempt to fool a model ($\model$) at inference time so that it produces an incorrect output. 
A common approach is to craft adversarial examples by adding small imperceptible perturbations to the input that change $\model$'s prediction~\cite{madry2018DeepLearningModels}.  

\item\label{poison} \textbf{Poisoning} manipulates the training process by injecting malicious records into the dataset.  
These records shift $\model$'s decision boundary and reduce utility.
Backdoor attacks~\cite{gu2019BadNetsEvaluatingBackdooring} are a special case where inputs containing a specific trigger are consistently misclassified.
Backdoors extend to text, where the trigger is an inserted word or phrase~\cite{cui2022UnifiedEvaluationTextual}.

\item\label{modelext} \textbf{Unauthorized model ownership} is the risk that an adversary derives a stolen model ($\modelstol$) that mimics the behavior of $\model$.  
$\modelstol$ may be an identical copy of $\model$ or obtained through model extraction, where it is trained on the outputs of $\model$~\cite{orekondy2019KnockoffNetsStealing}.  
\end{enumerate}

\begin{enumerate}[label=\textbf{P\arabic*},leftmargin=*,wide, labelindent=0pt, itemsep=0pt, topsep=0pt]
\itemsep0em 
\item\label{meminf} \textbf{Membership inference} aims to determine whether a specific record was part of $\model$'s training data ($\dtrain$).  
These attacks exploit differences in how $\model$ behaves on records from $\dtrain$ compared to those outside it, which can reveal sensitive information~\cite{carlini2022MembershipInferenceAttacks}.  

\item\label{attinf} \textbf{Attribute inference} seeks to infer the value of a hidden attribute that is not part of the input or $\dtrain$~\cite{aalmoes2024AlignmentGroupFairness}. 
When the attribute is sensitive, such as race or gender, these attacks can expose information that the data owner may not wish to disclose.

\item\label{datarecon} \textbf{Data reconstruction} recovers the average data record of a particular class in $\dtrain$ using the outputs of $\model$~\cite{fredrikson2015ModelInversionAttacks}.  
Some attacks may also recover individual data records~\cite{carlini2021ExtractingTrainingData}.  

\item\label{distinf} \textbf{Distribution inference} aims to infer statistical properties of $\dtrain$ (e.g., class-imbalance or demographic proportions)~\cite{suri2023DissectingDistributionInference}.  
A common approach is hypothesis testing, where the adversary trains shadow models on datasets with different candidate distributions and uses them to infer the distribution of $\dtrain$ reflected in $\model$. 
\end{enumerate}

\begin{enumerate}[label=\textbf{F},leftmargin=*,wide, labelindent=0pt, itemsep=0pt, topsep=0pt]
\item\label{disc} \textbf{Discriminatory behavior} occurs when $\model$ produces systematically different outcomes across demographic subgroups~\cite{dwork2012FairnessAwareness}.  
Such disparities may stem from bias in $\dtrain$ or the training process.  
\end{enumerate}

\subsection{Machine Learning Defenses}\label{sec:backdef}

We present corresponding defenses:
security risks (\ref{advtr},~\ref{outrem},~\ref{fngrprnt}, and~\ref{wm}),
privacy risks (\ref{dpsgd}) and
fairness risk (\ref{gpfair}).
\begin{enumerate}[label=\textbf{SD\arabic*},leftmargin=*,wide, labelindent=0pt, itemsep=0pt, topsep=0pt]
\itemsep0em 
\item\label{advtr} \textbf{Adversarial training} protects against~\ref{evasion} by training $\model$ with adversarial examples~\cite{madry2018DeepLearningModels}.  
It does so by minimizing the maximum loss over worst-case adversarial perturbations.  

\item\label{outrem} \textbf{Outlier removal} protects against~\ref{poison} by limiting the impact of malicious data.
Approaches include detecting and removing poisoned records from $\dtrain$~\cite{tran2018SpectralSignaturesBackdoor}, suppressing them during training~\cite{li2021AntiBackdoorLearningTraining}, or pruning $\model$ after training~\cite{wu2021AdversarialNeuronPruning}.
A complementary approach purifies inputs at inference by removing trigger content; for text, ONION deletes tokens whose removal a language model finds anomalous~\cite{qi2021ONIONSimpleEffective}.

\item\label{fngrprnt} \textbf{Fingerprinting} is a post-hoc defense against~\ref{modelext} that extracts a unique signature from both the victim and suspect models to distinguish them~\cite{maini2021DatasetInferenceOwnership}.  
Fingerprints can be derived from various model characteristics, such as transferable adversarial examples~\cite{lukas2021DeepNeuralNetwork} or decision boundaries~\cite{maini2021DatasetInferenceOwnership}.  

\item\label{wm} \textbf{Watermarking} is a defense against~\ref{modelext} that embeds marked data records into $\dtrain$ so they can later be extracted to verify ownership~\cite{adi2018TurningYourWeakness}.  
These watermarks are typically out-of-distribution records memorized by $\model$ and reliably recovered during verification~\cite{adi2018TurningYourWeakness}.  
\end{enumerate}

\begin{enumerate}[label=\textbf{PD\arabic*},leftmargin=*,wide, labelindent=0pt, itemsep=0pt, topsep=0pt]
\item\label{dpsgd} \textbf{Differential privacy}, specifically differentially private stochastic gradient descent (DP-SGD)~\cite{abadi2016DeepLearningDifferential}, protects against~\ref{meminf},~\ref{attinf}, and~\ref{datarecon}.
DP-SGD clips gradients and adds noise during training, ensuring $\model$'s outputs are nearly indistinguishable with or without any single record in $\dtrain$. To the best of our knowledge, no defenses provide theoretical guarantees against~\ref{distinf}.
\end{enumerate}

\begin{enumerate}[label=\textbf{FD\arabic*},leftmargin=*,wide, labelindent=0pt, itemsep=0pt, topsep=0pt]
\itemsep0em
\item\label{gpfair} \textbf{Adversarial debiasing} encourages $\model$ to behave similarly across demographic subgroups.  
A common approach is to use a GAN-style setup where an adversary acts as a discriminator, trying to predict subgroup membership from $\model$'s output~\cite{zhang2018MitigatingUnwantedBiases, louppe2017LearningPivotAdversarial}.  
\end{enumerate}

\subsection{Interactions among Defenses and Risks}\label{sec:interactions}

Most defenses are designed to address a single risk (as discussed in Section~\ref{sec:backdef}).  
In practice, a deployed model may be susceptible to several risks simultaneously.  
Practitioners must therefore understand (a) the conflicts that arise when combining defenses against multiple risks~\cite{szyller2023ConflictingInteractionsProtection}, and (b) the interactions of defenses with risks outside their original scope~\cite{duddu2024SoKUnintendedInteractions}.  

\noindent\textbf{Conflicts among defenses.} Defenses can be combined to address multiple risks simultaneously~\cite{duddu2025CombiningMachineLearning}.  
However, such combinations may introduce conflicts that reduce their effectiveness or degrade $\model$'s utility~\cite{szyller2023ConflictingInteractionsProtection}.  
These conflicts can be viewed as unintended interactions among defenses.  

\noindent\textbf{Unintended interactions among defenses and risks.} A defense designed for one risk may inadvertently increase or decrease susceptibility to others, an \emph{unintended interaction}~\cite{duddu2024SoKUnintendedInteractions, szyller2023ConflictingInteractionsProtection}.  
For example, adversarial training can increase susceptibility to privacy~\cite{song2019PrivacyRisksSecuring} and fairness risks~\cite{benz2021RobustnessMayBe}.  
Duddu et al.~\cite{duddu2024SoKUnintendedInteractions} identified overfitting and memorization as key drivers of such interactions, since defenses influence these factors and thereby affect susceptibility to risks whose attacks exploit them.  
Other factors shaping overfitting and memorization (e.g., model size, dataset size, distance to decision boundary) also influence unintended interactions, but a comprehensive evaluation is still missing.  
In this work, we focus on unintended interactions between defenses and risks.

\subsection{Other Libraries}

Among privacy-focused libraries, ML-Doctor~\cite{liu2022MLDoctorHolisticRisk} is the most comprehensive.  
It consolidates attacks such as model extraction, model inversion, membership inference, and attribute inference, benchmarks them, and studies how overfitting and data complexity affect privacy risks.  
ML Privacy Meter~\cite{murakonda2020MLPrivacyMeter}, in contrast, is limited to membership inference.  
Other tools offer implementations and benchmarks for differentially private training~\cite{yousefpour2021OpacusUserFriendlyDifferential,wei2023DPMLBenchHolisticEvaluation}.  

Several frameworks address the security of ML models.  
They focus on robustness by testing models against evasion and poisoning attacks, and often include adversarial training~\cite{papernot2016TechnicalReportCleverHans,ding2019AdvertorchV01Adversarial,nicolae2018AdversarialRobustnessToolbox,ling2019DEEPSECUniformPlatform}.
Adversarial Robustness Toolbox is the most comprehensive and also covers certain privacy risks~\cite{nicolae2018AdversarialRobustnessToolbox}.  
TROJANZOO,~\cite{pang2022TROJANZOOUnifiedHolistic} another security-focused library, primarily implements backdoor attacks.  

Finally, several libraries address fairness in ML.  
Some focus on model explanations, such as Captum~\cite{kokhlikyan2020CaptumUnifiedGeneric}, while others implement algorithmic fairness techniques~\cite{bellamy2018AIFairness360,weerts2023FairlearnAssessingImproving}.  

\section{Problem Statement}\label{sec:problem}

We aim to design a library that evaluates both intended and unintended interactions among ML defenses and risks.
This is challenging because risks rely on different datasets, threat models, and implementations, and evaluation practices are often flawed or inconsistent~\cite{carlini2022MembershipInferenceAttacks}.
Standardization across datasets, threat models, and evaluation protocols is therefore necessary to enable a systematic assessment of interactions.
A library must be:
\begin{enumerate}[leftmargin=*,label=\textbf{D\arabic*},itemjoin={,\xspace}] 
\item\label{comprehensive} \textbf{comprehensive}, covering state-of-the-art attacks, defenses, and metrics for different risks;
\item\label{consistent} \textbf{consistent}, offering a user-friendly API;
\item\label{extensible} \textbf{extensible}, allowing easy addition of new modules;
\item\label{applicable} \textbf{applicable}, enabling evaluation of unintended interactions among ML defenses and risks.
\end{enumerate}

\paragraph{Limitations of prior work.}
We surveyed 11 libraries, summarized in Table~\ref{tab:prior_work}, where \cmark denotes requirement is fully satisfied, \smark is partially satisfied, and \xmark is not satisfied. 
We discuss how each library fulfills the requirements.

\noindent\textit{\ref{comprehensive} Comprehensive.} 
Seven libraries partially satisfy this requirement (\smark), each within a specific category of risks.
The most comprehensive in their respective categories are ART~\cite{nicolae2018AdversarialRobustnessToolbox} for security, ML-Doctor~\cite{liu2022MLDoctorHolisticRisk} for privacy, and Fairlearn~\cite{weerts2023FairlearnAssessingImproving} for fairness.
However, no library includes modules spanning all three types of risk.

\noindent\textit{\ref{consistent} Consistent.}
Apart from DPMLBench~\cite{wei2023DPMLBenchHolisticEvaluation},
all listed libraries are internally consistent (\cmark),
providing easy-to-use APIs that can be plugged into ML pipelines.
However, because their designs diverge, comparing risks across different libraries remains difficult.

\pgfdeclarelayer{background}
\pgfdeclarelayer{foreground}
\pgfsetlayers{background,main,foreground}

\tikzstyle{property} = [rectangle,  minimum width=1.4cm, minimum height=0.5cm, text centered, draw=black, fill=blue!30]
\tikzstyle{defense} = [rectangle,  minimum width=1.4cm, minimum height=0.5cm, text centered, draw=black, fill=green!30]
\tikzstyle{risk} = [rectangle,  minimum width=1.4cm, minimum height=0.5cm, text centered, draw=black, fill=orange!30]
\tikzstyle{attack} = [rectangle, minimum width=1.4cm, minimum height=0.5cm, text centered, draw=black, fill=red!20]
\tikzstyle{metrics} = [rectangle,  minimum width=1cm, minimum height=0.5cm, text centered, draw=black, fill=yellow!30]

\begin{figure*}[t]
\centering
\begin{tikzpicture}
\node (amulet) [rectangle, draw=black] {\footnotesize \method};

\node (privacy) [property, below of=amulet, yshift=-0.4cm] {\footnotesize Privacy};
\node (security) [property, left of=privacy, xshift=-5cm] {\footnotesize Security};
\node (fairness) [property, right of=privacy, xshift=5cm] {\footnotesize Fairness};



\draw[->, ultra thick] (amulet.south) -- ([xshift=0.65cm]security.north);
\draw[->, ultra thick] (amulet.south) -- (privacy.north);
\draw[->, ultra thick] (amulet.south) -- ([xshift=-0.65cm]fairness.north);

\node (evasion) [risk, below of=security, yshift=-0.2cm, minimum width=2.2cm] {\footnotesize Evasion};
\node (poisoning) [risk, below of=evasion, yshift=0.3cm, minimum width=2.2cm] {\footnotesize Poisoning};
\node (modelext) [risk, below of=poisoning,align=center, yshift=0.1cm] {\footnotesize Unauthorized\\\footnotesize Model Ownership};


\node (meminf) [risk, below of=privacy, align=center, yshift=-0.2cm] {\footnotesize Membership Inference};
\node (attinf) [risk, below of=meminf, align=center, yshift=0.3cm,minimum width=2.8cm] {\footnotesize Attribute Inference};
\node (distinf) [risk, below of=attinf, align=center,yshift=0.3cm,minimum width=2.8cm] {\footnotesize Distribution Inference};
\node (datarecon) [risk, below of=distinf, align=center,yshift=0.3cm,minimum width=2.8cm] {\footnotesize Data Reconstruction};

\node (disc) [risk, below of=fairness,align=center, yshift=-0.4cm] {\footnotesize Discriminatory\\\footnotesize Behavior};


\begin{scope}[on background layer]
    \node (risks1) [fit=(evasion) (poisoning) (modelext), rounded corners, draw=black, dashed, inner sep=.2cm,fill= orange!15] {};
\end{scope}

\begin{scope}[on background layer]
    \node (risks2) [fit=(meminf) (attinf) (distinf) (datarecon), rounded corners, draw=black, dashed, inner sep=.2cm,fill= orange!15] {};
\end{scope}

\begin{scope}[on background layer]
    \node (risks3) [fit=(disc), rounded corners, draw=black, dashed, inner sep=.2cm,fill= orange!15] {};
\end{scope}

\draw[->, ultra thick] (security.south) -- (risks1.north);
\draw[->, ultra thick] (privacy.south) -- (risks2.north);
\draw[->, ultra thick] (fairness.south) -- (risks3.north);

\node (attacks1) [attack, below of=modelext, yshift=-0.7cm, xshift=-1.8cm] {\footnotesize Attack 1};
\node (attacksN) [attack, below of=attacks1, yshift=-0.3cm] {\footnotesize Attack N};

\begin{scope}[on background layer]
    \node (attacks) [fit=(attacks1) (attacksN), rounded corners, draw=black, dashed, inner sep=.05cm,fill= red!15] {};
\end{scope}

\draw[->, ultra thick] (risks1.south) -- (attacks.north);

\node (defense1) [defense, below of=modelext, yshift=-0.7cm] {\footnotesize Defense 1};
\node (defenseN) [defense, below of=defense1, yshift=-0.3cm] {\footnotesize Defense N};

\begin{scope}[on background layer]
    \node (defenses) [fit=(defense1) (defenseN), rounded corners, draw=black, dashed, inner sep=.05cm,fill= green!15] {};
\end{scope}

\draw[->, ultra thick] (risks1.south) -- (defenses.north);

\node (metric1) [metrics, below of=modelext, yshift=-0.7cm,xshift=1.8cm] {\footnotesize Metric 1};
\node (metricN) [metrics, below of=metric1, yshift=-0.3cm] {\footnotesize Metric N};

\begin{scope}[on background layer]
    \node (metric) [fit=(metric1) (metricN), rounded corners, draw=black, dashed, inner sep=.05cm,fill= yellow!15] {};
\end{scope}

\draw[->, ultra thick] (risks1.south) -- (metric.north);

\node (attacks1) [attack, right of=metric1, xshift=1.4cm] {\footnotesize Attack 1};
\node (attacksN) [attack, below of=attacks1, yshift=-0.3cm] {\footnotesize Attack N};

\begin{scope}[on background layer]
    \node (attacks) [fit=(attacks1) (attacksN), rounded corners, draw=black, dashed, inner sep=.05cm,fill= red!15] {};
\end{scope}

\draw[->, ultra thick] (risks2.south) -- (attacks.north);

\node (defense1) [defense, right of=attacks1,xshift=0.8cm] {\footnotesize Defense 1};
\node (defenseN) [defense, below of=defense1, yshift=-0.3cm] {\footnotesize Defense N};

\begin{scope}[on background layer]
    \node (defenses) [fit=(defense1) (defenseN), rounded corners, draw=black, dashed, inner sep=.05cm,fill= green!15] {};
\end{scope}

\draw[->, ultra thick] (risks2.south) -- (defenses.north);

\node (metric1) [metrics, right of=defense1, xshift=0.8cm] {\footnotesize Metric 1};
\node (metricN) [metrics, below of=metric1, yshift=-0.3cm] {\footnotesize Metric N};

\begin{scope}[on background layer]
    \node (metric) [fit=(metric1) (metricN), rounded corners, draw=black, dashed, inner sep=.05cm,fill= yellow!15] {};
\end{scope}

\draw[->, ultra thick] (risks2.south) -- (metric.north);

\node (attacks1) [attack, right of=metric1, xshift=1.4cm] {\footnotesize Attack 1};
\node (attacksN) [attack, below of=attacks1, yshift=-0.3cm] {\footnotesize Attack N};

\begin{scope}[on background layer]
    \node (attacks) [fit=(attacks1) (attacksN), rounded corners, draw=black, dashed, inner sep=.05cm,fill= red!15] {};
\end{scope}

\draw[->, ultra thick] (risks3.south) -- (attacks.north);

\node (defense1) [defense, right of=attacks1,xshift=0.8cm] {\footnotesize Defense 1};
\node (defenseN) [defense, below of=defense1, yshift=-0.3cm] {\footnotesize Defense N};

\begin{scope}[on background layer]
    \node (defenses) [fit=(defense1) (defenseN), rounded corners, draw=black, dashed, inner sep=.05cm,fill= green!15] {};
\end{scope}

\draw[->, ultra thick] (risks3.south) -- (defenses.north);

\node (metric1) [metrics, right of=defense1, xshift=0.8cm] {\footnotesize Metric 1};
\node (metricN) [metrics, below of=metric1, yshift=-0.3cm] {\footnotesize Metric N};

\begin{scope}[on background layer]
    \node (metric) [fit=(metric1) (metricN), rounded corners, draw=black, dashed, inner sep=.05cm,fill= yellow!15] {};
\end{scope}

\draw[->, ultra thick] (risks3.south) -- (metric.north);

\end{tikzpicture}
\caption{\underline{\textbf{Overview of \method{'s} design:}} Top level modules are categorized into different properties (\colorbox{blue!20}{blue}): security, privacy, and fairness. For each property, we include different risks which violate them (\colorbox{orange!20}{orange}). Under each risk, we include attacks which realize the risk (in \colorbox{red!20}{red}), defenses to protect against it (in \colorbox{green!20}{green}), and metrics to evaluate risk susceptibility (\colorbox{yellow!20}{yellow}).}
\label{fig:overview}
\Description{Hierarchical diagram of Amulet's modular design. The Amulet root branches into three properties (security, privacy, fairness); each property branches into the risks that violate it; and each risk connects to the attacks, defenses, and metrics used to evaluate risk susceptibility.}
\end{figure*}

\noindent\textit{\ref{extensible} Extensible.}
We consider a library extensible if users can easily contribute new attacks, defenses, or metrics.
In the absence of a contributing guide, the requirement is not met.
A minimal guide partially satisfies it (\smark), while a detailed guide with coding standards fully satisfies it (\cmark).
By this measure, five libraries are extensible, while ART~\cite{nicolae2018AdversarialRobustnessToolbox} only partially satisfies the requirement.

\noindent\textit{\ref{applicable} Applicable.}
Although several libraries are comprehensive, extensible, and consistent, none are applicable for evaluating unintended interactions across different risks (\xmark).
At best, users can examine intended interactions when both attacks and defenses are available.

\begin{takeaway}
Each surveyed library satisfies some requirements but not all four together, so studying interactions across risks stays piecemeal.
\method is designed to satisfy all four.
\end{takeaway}

\begin{table}[t]
    \centering
    \small
    \caption{\textbf{Summary of prior work.} \cmark, \smark, and \xmark denote whether the requirement is satisfied fully, partially, or not at all.}
    \begin{tabular}{l l c c c}
        \bottomrule

        \toprule
        \textbf{Library} & 
        \ref{comprehensive} & \ref{consistent}~ & \ref{extensible} & \ref{applicable} \\ 
        \bottomrule

        \toprule
        \textbf{CleverHans}~\cite{papernot2016TechnicalReportCleverHans} & \smark & \cmark & \cmark & \xmark \\ 
        \textbf{ART}~\cite{nicolae2018AdversarialRobustnessToolbox} & \smark & \cmark & \smark & \xmark \\ 
        \textbf{DEEPSEC}~\cite{ling2019DEEPSECUniformPlatform} & \smark & \cmark & \xmark & \xmark \\ 
        \textbf{TrojanZoo}~\cite{pang2022TROJANZOOUnifiedHolistic} & \xmark & \cmark & \xmark & \xmark \\
        \textbf{ML-Doctor}~\cite{liu2022MLDoctorHolisticRisk} & \smark & \cmark & \cmark & \xmark \\ 
        \textbf{Priv. Mtr.}~\cite{murakonda2020MLPrivacyMeter} & \smark & \cmark & \xmark & \xmark \\ 
        \textbf{DPMLBench}~\cite{wei2023DPMLBenchHolisticEvaluation} & \xmark & \smark & \xmark & \xmark \\ 
        \textbf{Opacus}~\cite{yousefpour2021OpacusUserFriendlyDifferential} & \xmark & \cmark & \xmark & \xmark \\
        \textbf{AIF360}~\cite{bellamy2018AIFairness360} & \smark & \cmark & \cmark & \xmark \\
        \textbf{Fairlearn}~\cite{weerts2023FairlearnAssessingImproving} & \smark & \cmark & \cmark & \xmark \\ 
        \textbf{Captum}~\cite{kokhlikyan2020CaptumUnifiedGeneric} & \xmark & \cmark & \cmark & \xmark \\
        \bottomrule

        \toprule
    \end{tabular}

    \label{tab:prior_work}
\end{table}
\section{Design of \method}\label{sec:design}
\noindent\textbf{Overview.}
\method is organized around \emph{risks} as the primary abstraction.
For each risk, we implement three pluggable module types: 
\emph{(i)} attacks which realize the risk,
\emph{(ii)} defenses which protect against the risk, and 
\emph{(iii)} metrics which evaluate susceptibility to the risk. 
This modular structure is illustrated in Figure~\ref{fig:overview} and ensures that new components can be added consistently across domains.

\noindent\textbf{Intended users.}
\method is designed to serve two groups: 
\begin{enumerate*}[label=\roman*),itemjoin={,\xspace}] 
    \item \emph{researchers}, who investigate new interactions, analyze how attack, defense, and metric variants affect them, and design defenses to mitigate or leverage these interactions
    \item \emph{practitioners}, who analyze trade-offs in deploying ML defenses specific to their setting.
\end{enumerate*}

\noindent\textbf{Scope.}
This design directly addresses the desiderata in Section~\ref{sec:problem}.
Below, we demonstrate how \method satisfies D1--D2 (comprehensive, consistent) by construction.
We evaluate D3 (extensibility) and D4 (applicability) empirically in Sections~\ref{sec:extensibility} and~\ref{sec:evaluation}.

\begin{table*}[t]
    \centering
    \small
    \caption{Attacks, defenses, and metrics currently implemented in \method.}
    \begin{tabular}{ l l l l }
        \bottomrule
        
        \toprule
        \textbf{Risk} & \textbf{Attack} & \textbf{Defense} & \textbf{Metric} \\ 
        \bottomrule
        
        \toprule
        \ref{evasion} & PGD~\cite{madry2018DeepLearningModels} & Adversarial Training~\cite{madry2018DeepLearningModels} & Robust Accuracy \\ \midrule
        \ref{poison} & BadNets~\cite{gu2019BadNetsEvaluatingBackdooring}, TextBadNets~\cite{cui2022UnifiedEvaluationTextual} & Outlier Removal, ONION~\cite{qi2021ONIONSimpleEffective} & Poisoned Data Accuracy, Attack Success Rate \\ \midrule
        \ref{modelext} & Tram\`{e}r et al.~\cite{tramer2016StealingMachineLearning} & Fingerprinting~\cite{maini2021DatasetInferenceOwnership}, Watermarking~\cite{adi2018TurningYourWeakness} & Accuracy, FPR, FNR \\ \midrule
        \ref{meminf} & LiRA~\cite{carlini2022MembershipInferenceAttacks} & DP-SGD~\cite{abadi2016DeepLearningDifferential} & ROC AUC, Balanced Accuracy, TPR@$1\%$FPR \\ \midrule
        \ref{attinf} & Aalmoes et al.~\cite{aalmoes2024AlignmentGroupFairness} & - & Balanced Accuracy, ROC AUC \\ \midrule
        \ref{distinf} & Suri et al.~\cite{suri2023DissectingDistributionInference} & - & Attack Accuracy \\ \midrule
        \ref{datarecon} & Fredrikson et al.~\cite{fredrikson2015ModelInversionAttacks} & DP-SGD~\cite{abadi2016DeepLearningDifferential} & MSE, SSIM \\ \midrule 
        \multirow{2}{*}{\ref{disc}} & \multirow{2}{*}{-} & \multirow{2}{*}{Adversarial Debiasing} & Demographic Parity, True Positive Parity, False Positive Parity \\
        & & & Equalized Odds, P-Rule \\
        \bottomrule
        
        \toprule
    \end{tabular}
    \label{tab:modules}
\end{table*}

\subsection{Comprehensive}

\method spans eight risks across security, privacy, and fairness, with representative attacks, defenses, and metrics summarized in Table~\ref{tab:modules}.
We deliberately favor breadth by implementing widely used and generalizable modules across all risks, rather than depth in any single risk category. 
This approach ensures coverage of diverse risk types while leaving room for the open-source community to expand each category. 
Our emphasis on extensibility further enables contributions of new attacks and defenses.

\method also includes components for datasets, models, and utilities that enable end-to-end evaluation.

\noindent\textbf{Datasets.}
\method provides five widely used datasets spanning vision and tabular domains:
\fmnist~\cite{xiao2017FashionMNISTNovelImage}, \cifar, \census~\cite{kohavi1996CensusIncome}, \lfw~\cite{melis2019ExploitingUnintendedFeature}, and \celeba~\cite{liu2015DeepLearningFace}.
These datasets cover small grayscale images, natural images, demographic records, and large-scale face attributes,
supporting evaluation across security, privacy, and fairness.
Section~\ref{sec:extensibility} extends this set to text with \sstwo~\cite{socher2013RecursiveDeepModels}.

\noindent\textbf{Model configurations.}
\method includes commonly used architectures for quick testing, such as VGG~\cite{liu2015VeryDeepConvolutional} and ResNet~\cite{he2016DeepResidualLearning}.
Section~\ref{sec:extensibility} adds an LLM, a LoRA-adapted \llama~\cite{llamateammetaai2024Llama3Herd}.

\subsection{Consistent}
Each risk in \method follows a uniform interface: attacks and defenses take a PyTorch \texttt{nn.Module} and hyperparameters as input (except poisoning attacks, which operate on datasets), and their outputs serve as inputs to the corresponding metrics.  
This consistent design allows \method to integrate into any PyTorch pipeline.  

Furthermore, \method offers helper functions for setting up an end-to-end pipeline, including data loading, model initialization, training, and evaluation.  
Each helper function, along with every attack and defense, provides default values that yield reasonable results when used with a VGG architecture on \celeba or \cifar.  

Attacks, defenses, and metrics follow a consistent layout:
\begingroup
\setlength{\topsep}{0pt}
\setlength{\partopsep}{0pt}
\begin{quote}
  \texttt{amulet.<risk>.attacks.<attack>} \\
  \texttt{amulet.<risk>.defenses.<defense>} \\
  \texttt{amulet.<risk>.metrics.<metric>}
\end{quote}
\endgroup

\begin{figure*}[htb]
\centering
\begingroup
\fvset{frame=lines,framesep=1.5mm,baselinestretch=1.0,fontsize=\footnotesize}%
\input{pygmented/fig_code-1.tex}%
\endgroup

\caption{Evasion, adversarial training, and model extraction using \method. 
Each attack or defense is instantiated as an object; attacks are run with \texttt{attack()}, while defenses use \texttt{train\_robust()}, \texttt{train\_fair()}, or \texttt{train\_private()} depending on type.}
\label{fig:code}
\Description{Python code listing showing how to run an evasion attack, an adversarial-training defense, and a model-extraction attack with Amulet. Each attack or defense is instantiated as an object and invoked through a small number of API calls.}
\end{figure*}

Figure~\ref{fig:code} illustrates example usage, assuming standard variables common to most ML pipelines.  
Each algorithm is instantiated as an object and then executed via a function call.  
More detailed examples are available in the \texttt{examples} directory of our repository.\footnote{\url{https://github.com/ssg-research/amulet}}

\method also provides an \texttt{AmuletDataset} class.
Converting a dataset into this format makes it compatible with any module in \method.
All modules support image-based (two-dimensional), tabular (one-dimensional), and text datasets.
Section~\ref{sec:extensibility} describes how we added the text modality and what it cost.

\begin{takeaway}
\method provides a single uniform interface for evaluating interactions across security, privacy, and fairness.
\end{takeaway}

\section{Evaluating Extensibility (D3)}\label{sec:extensibility}

The majority of the work in this space is conducted using outdated albeit standard benchmarks and datasets, centered around image classification.
On the other hand, attacks and defenses for LLMs are a rapidly changing field that lacks established baselines.
In this section, we show that \method is easily extended to LLMs.
We then evaluate an intended and an unintended interaction on the new modality with a LoRA-adapted \llama victim: ONION defending against a textual backdoor attack, and DP-SGD's effect on susceptibility to the same attack.

\noindent\textbf{\method modularity.}
\method organizes modules by risk. A new attack, defense, or metric is implemented under the namespace of the risk it targets, and the modules already present in that risk are unchanged (Figure~\ref{fig:overview}).
Each risk defines an abstract base class that fixes the entry point its defenses expose: defenses against~\ref{poison} implement a robust-training method, and defenses against~\ref{meminf} implement a private-training method.
A conformance test asserts that every defense exposes the entry point defined for its risk, and fails the build otherwise.
New datasets enter through the \texttt{AmuletDataset} class, and every helper function accepts configurable parameters.
\method integrates into an existing PyTorch pipelines.

\method provides a contribution guide and uses standard, modern Python tooling
\footnote{
\url{https://docs.astral.sh/uv/},
\url{https://docs.astral.sh/ruff/},
\url{https://microsoft.github.io/pyright/},
\url{https://pre-commit.com/}
}
:
\texttt{uv} for development environments,
\texttt{ruff} and
\texttt{pyright} for code quality checks, and
\texttt{pre-commit}
hooks that run those checks before each commit.

\noindent\textbf{Extending \method to LLMs.}
Adding a text modality required only three types of modules that slot into the existing framework.
First, \texttt{TextTensorDataset} represents a tokenized text dataset, extending the \texttt{AmuletDataset} class to admit text alongside images and tabular records.
Second, \texttt{HFCausalLM} wraps a large language model with low-rank adapters~\cite{hu2021LoRALowRankAdaptation} and a linear classification head for text classification.
Third, we added \texttt{TextBadNets}~\cite{cui2022UnifiedEvaluationTextual}, which implements backdoor poisoning for text, and \texttt{ONION}~(\ref{outrem}), a perplexity-based input purification defense against~\ref{poison}~\cite{qi2021ONIONSimpleEffective}.
Each module implements the abstract base class for its risk, so existing metrics for~\ref{poison} consume their outputs without modification.
\textbf{The extension required no changes to any existing risk, attack, defense, or metric.}

For the unintended interaction study, we reused DP-SGD~\cite{abadi2016DeepLearningDifferential}, already implemented, without modification.
DP-SGD had been developed and tested only on vision and tabular data, and it now runs on a three-billion-parameter LLM.

\noindent\textbf{Experimental Setup.}
We use \sstwo~\cite{socher2013RecursiveDeepModels} for both studies and measure the effect of ONION and DP-SGD on~\ref{poison}.
\sstwo provides $67{,}349$ training records, and we use its $872$-record validation split as the test set, since the labels of the official test split are withheld.
The victim is \llama~\cite{llamateammetaai2024Llama3Herd}, adapted with LoRA at rank $8$ and a linear classification head over the pooled final-token representation, trained in full precision.
\texttt{TextBadNets} inserts the trigger \texttt{cf} at a random position in a poisoned record and assigns that record the target label.
Random placement follows the protocol of the original evaluation and prevents ONION from identifying the trigger by position alone.

Each study compares three victims.
$\modelstd$ is trained on the clean dataset as the baseline.
$\modelpois$ trains on the poisoned dataset with no defense.
$\modeldef$ trains on the same poisoned dataset with the defense applied.
We report clean test set accuracy ($Acc_{te}$), and attack success rate ($ASR$), the fraction of triggered test inputs that the victim assigns to the target class.
ONION purifies both the training data and the inputs at inference, so we evaluate $\modeldef$ on purified inputs.
DP-SGD applies during training only, so we evaluate its victim on unmodified inputs.
We sweep the poison rate from $0.01\%$ to $5\%$, corresponding to $6$ and $3{,}367$ records.
Each configuration is repeated five times.

\begin{table}[htb]
    \setlength\tabcolsep{3pt}
    \centering
    \small
    \caption{Poisoning~(\ref{poison}) versus ONION on \sstwo with a LoRA-adapted
    \llama victim. ONION is added to the poisoned model $\modelpois$; the clean
    baseline $\modelstd$ reaches $Acc_{te}=95.76\pm0.17$. ONION drives $ASR$ down at
    every poison rate.}%
    \label{tab:textbadnets_onion}
    \begin{tabular}{ c c c c c }
        \bottomrule

        \toprule
        \multirow{2}{*}{\textbf{Poisoned}} & \multicolumn{2}{c}{\textbf{Undefended} ($\modelpois$)} & \multicolumn{2}{c}{\textbf{Defended} ($\modeldef$)} \\
        & $Acc_{te}$ & $ASR$ & $Acc_{te}$ & $ASR$ \\
        \midrule
        0.01\% & 95.39~$\pm$~0.27 & 19.21~$\pm$~5.68 & 93.51~$\pm$~0.17 & 7.85~$\pm$~0.81 \\
        0.1\% & 95.48~$\pm$~0.37 & 98.27~$\pm$~1.16 & 94.22~$\pm$~0.17 & 8.04~$\pm$~0.72 \\
        1\% & 95.60~$\pm$~0.30 & 99.95~$\pm$~0.05 & 93.83~$\pm$~0.31 & 8.08~$\pm$~1.06 \\
        2\% & 95.57~$\pm$~0.17 & 99.95~$\pm$~0.05 & 93.94~$\pm$~0.48 & 6.78~$\pm$~0.60 \\
        5\% & 95.00~$\pm$~0.53 & 100.00~$\pm$~0.00 & 93.17~$\pm$~0.39 & 8.32~$\pm$~1.50 \\
        \bottomrule

        \toprule
    \end{tabular}
\end{table}

\begin{table}[htb]
    \setlength\tabcolsep{3pt}
    \centering
    \small
    \caption{Poisoning~(\ref{poison}) versus DP-SGD on \sstwo with a LoRA-adapted
    \llama victim, at privacy budgets $\epsilon\in\{1,8\}$. The clean baseline
    $\modelstd$ reaches $Acc_{te}=95.76\pm0.17$. DP-SGD suppresses the backdoor only at
    small poison rates.}%
    \label{tab:textbadnets_dpsgd}
    \begin{tabular}{ c c c c c c }
        \bottomrule

        \toprule
        \multirow{2}{*}{\textbf{Poisoned}} & \multirow{2}{*}{$\boldsymbol{\epsilon}$} & \multicolumn{2}{c}{\textbf{Undefended} ($\modelpois$)} & \multicolumn{2}{c}{\textbf{Defended} ($\modeldef$)} \\
        & & $Acc_{te}$ & $ASR$ & $Acc_{te}$ & $ASR$ \\
        \midrule
        \multirow{2}{*}{0.01\%} & 1 & \multirow{2}{*}{95.94~$\pm$~0.11} & \multirow{2}{*}{17.43~$\pm$~7.97} & 90.34~$\pm$~0.44 & 1.26~$\pm$~0.42 \\
         & 8 &  &  & 94.38~$\pm$~0.10 & 2.43~$\pm$~0.50 \\
        \midrule
        \multirow{2}{*}{0.1\%} & 1 & \multirow{2}{*}{95.87~$\pm$~0.13} & \multirow{2}{*}{98.93~$\pm$~0.96} & 90.09~$\pm$~0.36 & 1.96~$\pm$~0.60 \\
         & 8 &  &  & 94.22~$\pm$~0.19 & 3.60~$\pm$~0.81 \\
        \midrule
        \multirow{2}{*}{1\%} & 1 & \multirow{2}{*}{96.15~$\pm$~0.20} & \multirow{2}{*}{99.95~$\pm$~0.05} & 89.54~$\pm$~0.68 & 50.19~$\pm$~4.81 \\
         & 8 &  &  & 94.08~$\pm$~0.12 & 54.67~$\pm$~4.39 \\
        \midrule
        \multirow{2}{*}{5\%} & 1 & \multirow{2}{*}{95.87~$\pm$~0.16} & \multirow{2}{*}{100.00~$\pm$~0.00} & 88.12~$\pm$~0.71 & 98.55~$\pm$~0.19 \\
         & 8 &  &  & 93.81~$\pm$~0.20 & 97.29~$\pm$~0.88 \\
        \bottomrule

        \toprule
    \end{tabular}
\end{table}

\subsection{Intended Interaction:~\ref{poison} (Poisoning) and~\ref{outrem} (ONION)}

Results for the two studies are shown in Tables~\ref{tab:textbadnets_onion} and~\ref{tab:textbadnets_dpsgd}.
The attack reaches $98.3\%$ $ASR$ at $67$ poisoned records and $100\%$ above that, while $Acc_{te}$ stays within a point of the $95.76\%$ baseline.
ONION reduces $ASR$ to $6.8$--$8.3\%$ at a cost of roughly two points of $Acc_{te}$, and removed the trigger from every triggered input in every run.
The original evaluation of ONION on \sstwo reports near-total attack success without a defense, and single to low double digit $ASR$ under the defense at a small accuracy cost~\cite{qi2021ONIONSimpleEffective}.
Our results on \llama are consistent with these values.

The sweep also shows two effects that the original evaluation does not report.
First, $ASR$ under ONION holds near $8\%$ across a $560$-fold change in the poison count.
Once ONION removes the trigger, the poison count no longer affects the outcome.
Second, the attack has a minimum scale.
At $67$ poisoned records out of $67{,}349$ it succeeds nearly every time, while at $6$ records mean $ASR$ falls to $19.2\%$ with a standard error of $5.7$, and per-seed values span $2.1\%$ to $33.9\%$.
Backdooring an LLM of this size therefore takes very few poisoned records, and the attack is unreliable below that point.

\subsection{Unintended Interaction:~\ref{poison} (Poisoning) and~\ref{dpsgd} (DP-SGD)}

DP-SGD~(\ref{dpsgd}) protects against~\ref{meminf} and is not designed to protect against~\ref{poison}.
It nonetheless suppresses the backdoor while the poison count is small.
At $67$ poisoned records it reduces $ASR$ from $98.9\%$ to $2.0\%$ at $\epsilon=1$ and $3.6\%$ at $\epsilon=8$.
At $673$ records the reduction is partial, to $50.2\%$ and $54.7\%$.
At $3{,}367$ records it disappears: the privately trained victim reaches $98.6\%$ and $97.3\%$ $ASR$, leaving the attack as effective as it is against an undefended victim.
The tighter budget suppresses more at every poison count where the defense has an effect.
At $6$ poisoned records the attack is unreliable without any defense, reaching $17.4\%$ $ASR$, which reproduces the minimum scale the ONION sweep shows.

This pattern follows the group-privacy bound on poisoning a private learner.
An $\epsilon$-private learner forces an adversary to modify at least $\frac{1}{\epsilon}\log\tau$ records to reduce the attack cost by a factor of $\tau$, so the protection holds for small poison counts and decays as that count grows~\cite{ma2019DataPoisoningDifferentiallyPrivate}.
DP-SGD bounds the influence that any single training record exerts on the trained model, and its protection for a group of records weakens as that group grows.
A backdoor planted with $67$ records is a small set of outliers that clipping and noise suppress.
A backdoor planted with $3{,}367$ records forms a subpopulation whose gradient direction many records share, and a bound on individual influence does not constrain it.

Private training costs at most $2.0$ points of $Acc_{te}$ at $\epsilon=8$ and $7.6$ points at $\epsilon=1$, measured against the $95.76\%$ baseline.
At $67$ poisoned records this buys a $95$-point reduction in $ASR$, and at $3{,}367$ records it buys none.
Measuring this interaction at a single poison rate would mischaracterize the defense in either direction.

The bound certifies protection for small poison counts, but empirical studies of DP-SGD reach conclusions that depend on the setting.
Measured against a non-private baseline, private models can be more susceptible than undefended ones, because the accuracy that privacy removes and the accuracy that poisoning removes fall on the same records~\cite{jagielski2021DoesDifferentialPrivacy}.
A separate analysis attributes the poisoning benefit of DP-SGD to gradient clipping alone, with the useful configurations sitting at privacy budgets too loose to certify~\cite{hong2020EffectivenessMitigatingData}.
These studies evaluate convex models or small networks, and none reports a backdoor on an LLM.
Our result adds a point they do not cover: on a three-billion-parameter victim, DP-SGD suppresses the backdoor at $\epsilon=1$ when the poison count is small, a meaningful privacy budget, and costs under eight points of $Acc_{te}$.
The sign and size of this interaction depend on the model, the attack, and the privacy budget, and cannot be read off from any single study.

\begin{takeaway}
Adding a text modality required only three modules that slot into the API\@.
The implemented privacy defense applies directly to LLMs -- showing \method's extensibility.
\end{takeaway}

\section{Evaluating Applicability (D4)}\label{sec:evaluation}
\begin{table*}[htb]
    \centering
    \small
    \caption{Baseline evaluation of each risk in \method on \celeba, reproducing prior attacks to validate the implementations. $\modelstd$, $\modelpois$, and $\modelstol$ are the baseline, poisoned, and stolen models. $^{*}$Membership inference uses an intentionally overfit ResNet-18.}
    \label{tab:attack_results}
    \begin{tabularx}{\textwidth}{l l X X X X}
        \toprule
        \multirow{2}{*}{\textbf{Attack}} & \textbf{Model Architecture} & \textbf{VGG11} & \textbf{VGG13} & \textbf{VGG16} & \textbf{VGG19} \\
            & $Acc_{te}$ & 91.21~$\pm$~0.05 & 91.46~$\pm$~0.04 & 91.38~$\pm$~0.03 & 91.41~$\pm$~0.03 \\
        \midrule

        \ref{evasion}~(Evasion) & $Acc_{rob}$ & 10.08~$\pm$~0.13 & 8.60~$\pm$~0.05 & 8.65~$\pm$~0.02 & 8.67~$\pm$~0.04 \\
        \midrule

        \multirow{3}{*}{\ref{poison}~(Poisoning)} 
            &  $Acc_{pois}$ ($\modelstd$) & 8.13~$\pm$~0.30 & 8.27~$\pm$~0.34 & 8.17~$\pm$~0.19 & 7.57~$\pm$~0.31 \\
            & $Acc_{te}$ ($\modelpois$) & 91.16~$\pm$~0.03 & 91.48~$\pm$~0.04 & 91.39~$\pm$~0.03 & 91.37~$\pm$~0.08 \\
            & $Acc_{pois}$ ($\modelpois$) & 95.70~$\pm$~0.21 & 96.70~$\pm$~0.10 & 96.62~$\pm$~0.18 & 96.75~$\pm$~0.18 \\
        \midrule

        \multirow{3}{*}{\ref{modelext}~(Unauthorized Model Ownership)} 
            &  $Acc_{te}$ ($\modelstol$) & 91.02~$\pm$~0.09 & 91.32~$\pm$~0.07 & 91.42~$\pm$~0.08 & 91.39~$\pm$~0.06 \\
            & $Fid$ ($\modelstol$) & 94.64~$\pm$~0.16 & 94.75~$\pm$~0.09 & 94.74~$\pm$~0.11 & 94.84~$\pm$~0.08 \\
            & $Fid_{cor}$ ($\modelstol$) & 93.14~$\pm$~0.06 & 93.49~$\pm$~0.03 & 93.51~$\pm$~0.02 & 93.41~$\pm$~0.04 \\
        \midrule

        \multirow{8}{*}{\ref{meminf}~(Membership Inference)$^{*}$}
            & $Acc_{tr}$ & 94.20~$\pm$~1.33 & - & - & - \\
            & $Acc_{te}$ & 76.81~$\pm$~0.48 & - & - & - \\
            & Offline $Acc_{bal}$ & 50.41~$\pm$~00.13 & - & - & - \\
            & Offline $AUC$ & 0.49~$\pm$~0.00 & - & - & - \\
            & Offline TPR@1\%FPR & 0.83~$\pm$~0.11 & - & - & - \\
            & Online $Acc_{bal}$ & 60.67~$\pm$~00.70 & - & - & - \\
            & Online $AUC$ & 0.64~$\pm$~0.01 & - & - & - \\
            & Online TPR@1\%FPR & 2.70~$\pm$~0.25 & - & - & - \\
        \midrule
        
        \multirow{2}{*}{\ref{attinf}~(Attribute Inference)} 
            & $Acc_{bal}$ & 57.17~$\pm$~0.15 & 57.12~$\pm$~0.12 & 57.27~$\pm$~0.11 & 56.90~$\pm$~0.13 \\
            & $AUC$ & 0.59~$\pm$~0.00 & 0.59~$\pm$~0.00 & 0.59~$\pm$~0.00 & 0.58~$\pm$~0.00 \\
        \midrule
        
        \multirow{3}{*}{\ref{datarecon}~(Data Reconstruction)} 
            & $MSE_{avg}$ & 0.20~$\pm$~0.00 & 0.20~$\pm$~0.00 & 0.20~$\pm$~0.00 & 0.20~$\pm$~0.000 \\
            & $MSE_0$ & 0.19~$\pm$~0.00 & 0.19~$\pm$~0.00 & 0.19~$\pm$~0.00 & 0.19~$\pm$~0.00 \\
            & $MSE_1$ & 0.20~$\pm$~0.00 & 0.20~$\pm$~0.00 & 0.20~$\pm$~0.00 & 0.20~$\pm$~0.00 \\
        \bottomrule
    \end{tabularx}
\end{table*}

We first train a \celeba baseline to validate the correctness of the implementations and establish reference points for future comparisons. 
We train VGG11/13/16/19 models for 100 epochs as our main architectures using cross-entropy loss with the Adam optimizer unless otherwise specified. 
Each risk is then evaluated with its corresponding attack modules from Table~\ref{tab:modules}, and all experiments are repeated ten times with mean and standard error reported. 
We focus on \celeba since it combines image data with sensitive attributes, allowing us to assess multiple risks using one dataset. 
Next, we study three unintended interactions -- previously unexplored or only partially explained.

\subsection{Baseline Results}
Baseline results are shown in Table~\ref{tab:attack_results}. 
We evaluate each algorithm using \method{'s} metrics and compare to the original papers when possible. 
Overall, our reproductions closely match prior work, with exceptions noted below.

\noindent\textbf{\ref{evasion} (Evasion).}
Prior work evaluates PGD on \cifar and reports robust accuracy ($Acc_{rob}$) of $0.8$--$3.5\%$ for a ResNet model~\cite{madry2018DeepLearningModels}. 
With a comparable setup using \cifar, we obtained $Acc_{rob}=3.8\%$, consistent with these results. 
On \celeba, however, $Acc_{rob}$ is higher ($8.6$--$10\%$), making the attack less effective. 
We suspect this difference arises because \celeba ($\approx202{,}000$ images) is much larger than \cifar ($60{,}000$ images), and larger datasets typically yield more robust models~\cite{schmidt2018AdversariallyRobustGeneralization}.

\noindent\textbf{\ref{poison} (Poisoning).}
The original work evaluated BadNets on MNIST and traffic signs~\cite{gu2019BadNetsEvaluatingBackdooring}. 
Since MNIST is now considered too simple for robust evaluation, we reproduced it only for reference. 
Using a two-layer CNN, we get a poisoned data accuracy ($Acc_{pois}$) of $99\%$. 

On \celeba, the $\modelpois$ achieves $\approx96\%$ $Acc_{pois}$, demonstrating that the backdoor was successfully learned and the attack is effective.
In contrast, $\modelstd$ reaches only $\approx8\%$, since it never learned the trigger--target association.

\noindent\textbf{\ref{modelext} (Unauthorized Model Ownership).}
Prior work evaluated model extraction on \celeba with VGG19, combining three attributes into an 8-class classification problem~\cite{liu2022MLDoctorHolisticRisk}. 
They reported a best fidelity of $\approx81\%$ between $\modelstol$ and $\modelstd$. 

In contrast, \method achieves $\approx94\%$ fidelity on \celeba for the binary task, likely because the victim models in prior work were more overfit (Table 2 in~\cite{liu2022MLDoctorHolisticRisk}). 
We also report \emph{correct fidelity} ($Fid_{cor}$) for completeness, which measures agreement conditioned on correctness, i.e., when models agree, how often are they also correct. 

\noindent\textbf{\ref{meminf} (Membership Inference).}
To evaluate~\ref{meminf}, \method reports TPR at a fixed FPR of $1\%$, as recommended in prior work~\cite{carlini2022MembershipInferenceAttacks}. 
We also report balanced accuracy (best threshold from the ROC curve) and ROC AUC. 
Prior work evaluated this attack on \cifar with a ResNet model, using 256 shadow models, and achieved a TPR of $2.2\%$ at $0.001\%$ FPR, $8.4\%$ TPR at $0.1\%$ FPR, and a balanced accuracy of $63.8\%$~\cite{carlini2022MembershipInferenceAttacks}.

To reproduce comparable results, we trained an overfit ResNet-18 model, as shown in Table~\ref{tab:attack_results}. 
In contrast, a well-trained \celeba model is not susceptible to this attack.
This is because membership inference is more successful against overfit models~\cite{carlini2022MembershipInferenceAttacks}.

\noindent\textbf{\ref{attinf} (Attribute Inference).}
Performance is measured using balanced accuracy ($Acc_{bal}$) and ROC AUC. 
Prior work reported $53$--$62\%$ $Acc_{bal}$ across different datasets~\cite{aalmoes2024AlignmentGroupFairness}. 
On \celeba, we obtain $\approx57\%$ accuracy (Table~\ref{tab:attack_results}).
In addition, our preliminary tests on \census achieve $\approx65\%$ (Race) and $\approx59\%$ (Sex), closely matching the original results of $\approx56\%$ and $\approx58\%$. 
Overall, this suggests our results are consistent and reasonable.

\noindent\textbf{\ref{datarecon} (Data Reconstruction).}
Prior work applied the same reconstruction algorithm using a similar architecture on \celeba but with an 8-class setup (vs.\ our binary task) and reported a lower MSE ($\approx0.1$ vs. $\approx0.2$)~\cite{liu2022MLDoctorHolisticRisk}. 
Because the attack recovers an \emph{average} data record per class, having more classes reduces intra-class variability and typically yields lower MSE, explaining the discrepancy.

\begin{table*}[!htb]
    \setlength\tabcolsep{2.5pt}
    \small
    \centering
    \caption{Interaction between~\ref{advtr} Adversarial Training and~\ref{attinf} Attribute Inference on \census and \lfw. Adversarial training does not consistently change susceptibility to~\ref{attinf}.}
    \label{tab:attinf_advtr}
    \begin{tabular}{ c c c c c c c c }
        \bottomrule

        \toprule
        \multicolumn{8}{c}{\textbf{\census}}\\
            & $Acc_{te}$ & $Acc_{rob}$ ($\modelstd$) & $Acc_{rob}$ ($\modeldef$) & $Acc_{bal}$ (\race) & $AUC$ (\race) & $Acc_{bal}$ (\sex) & $AUC$ (\sex) \\
        \midrule
          Baseline ($\modelstd$) & 81.99~$\pm$~0.57 & - & - & 58.58~$\pm$~0.88 & 0.61~$\pm$~0.01 & 65.06~$\pm$~1.21 & 0.69~$\pm$~0.01 \\
          $\epsilon_{rob}$ = 0.01 & 82.53~$\pm$~0.44 & 73.87~$\pm$~1.18 & 81.15~$\pm$~0.68 & 59.64~$\pm$~0.53 & 0.62~$\pm$~0.01 & 65.88~$\pm$~0.80 & 0.70~$\pm$~0.01 \\
          $\epsilon_{rob}$ = 0.03 & 83.00~$\pm$~0.26 & 44.25~$\pm$~4.27 & 80.10~$\pm$~0.79 & 60.23~$\pm$~0.63 & 0.63~$\pm$~0.01 & 66.23~$\pm$~0.58 & 0.70~$\pm$~0.00 \\
          $\epsilon_{rob}$ = 0.06 & 83.40~$\pm$~0.36 & 22.70~$\pm$~1.95 & 80.03~$\pm$~0.64 & 60.28~$\pm$~0.31 & 0.63~$\pm$~0.00 & 66.78~$\pm$~1.09 & 0.71~$\pm$~0.01 \\
          $\epsilon_{rob}$ = 0.1 & 83.70~$\pm$~0.24 & 20.89~$\pm$~0.88 & 76.03~$\pm$~2.58 & 60.60~$\pm$~0.59 & 0.64~$\pm$~0.01 & 67.44~$\pm$~1.07 & 0.72~$\pm$~0.01 \\
        \midrule
        \multicolumn{8}{c}{\textbf{\lfw}}\\
            & $Acc_{te}$ & $Acc_{rob}$ ($\modelstd$) & $Acc_{rob}$ ($\modeldef$) & $Acc_{bal}$ (\race) & $AUC$ (\race) & $Acc_{bal}$ (\sex) & $AUC$ (\sex) \\
        \midrule
          Baseline ($\modelstd$) & 82.84~$\pm$~0.56 & - & - & 66.80~$\pm$~4.07 & 0.73~$\pm$~0.05 & 76.59~$\pm$~1.00 & 0.84~$\pm$~0.01 \\
          $\epsilon_{rob}$ = 0.01 & 81.36~$\pm$~0.67 & 54.31~$\pm$~1.48 & 75.09~$\pm$~0.95 & 68.41~$\pm$~6.43 & 0.74~$\pm$~0.08 & 76.16~$\pm$~0.74 & 0.84~$\pm$~0.01 \\
          $\epsilon_{rob}$ = 0.03 & 79.57~$\pm$~0.59 & 18.66~$\pm$~0.48 & 50.22~$\pm$~2.39 & 68.64~$\pm$~3.65 & 0.76~$\pm$~0.05 & 75.22~$\pm$~0.91 & 0.82~$\pm$~0.02 \\
          $\epsilon_{rob}$ = 0.06 & 78.47~$\pm$~0.66 & 17.20~$\pm$~0.55 & 33.28~$\pm$~2.58 & 72.37~$\pm$~2.79 & 0.79~$\pm$~0.04 & 75.06~$\pm$~1.30 & 0.82~$\pm$~0.01 \\
          $\epsilon_{rob}$ = 0.1 & 81.29~$\pm$~0.92 & 17.15~$\pm$~0.57 & 19.25~$\pm$~0.85 & 67.79~$\pm$~5.38 & 0.74~$\pm$~0.05 & 75.78~$\pm$~1.42 & 0.82~$\pm$~0.01 \\
        \bottomrule
        
        \toprule
    \end{tabular}
\end{table*}

\subsection{Unintended Interaction}
We now use \method to identify the nature of different unexplored unintended interactions. We focus on 
\begin{enumerate*}[label=\roman*),itemjoin={,\xspace}] 
\item \emph{unexplored interactions} and \item \emph{partially explored interactions} where the nature of the interaction is known, but the underlying factors that influence it are unclear.
\end{enumerate*}

We examine three such interactions: the effect of adversarial training on attribute inference and on unauthorized model ownership, and of outlier removal on unauthorized model ownership.

\noindent\textbf{\ref{advtr} (Adversarial Training) and~\ref{attinf} (Attribute Inference).} 
We use \census and \lfw to study how adversarial training (\ref{advtr}) influences attribute inference (\ref{attinf}), summarized in Table~\ref{tab:attinf_advtr}.  
In this experiment, we reserve 50\% of the training data for the adversary, reducing $Acc_{te}$ slightly.
Unless otherwise stated, we regard changes smaller than 5 p.p. or within one standard error as not significant, reflecting normal variation across runs.

On \census, $Acc_{bal}$ for \race rises only slightly, from $\approx58\%$ at baseline to $\approx60\%$ at $\epsilon_{rob}=0.01$, after which it remains stable.  
For the \sex attribute, $Acc_{bal}$ grows gradually from $\approx65\%$ to $\approx67\%$ at $\epsilon_{rob}=0.1$, with the $AUC$ following the same pattern.  
However, in both attributes, the overall change is only about 2 p.p., indicating that the effect of adversarial training on \census is negligible.

For \lfw, the change in $Acc_{bal}$ is more significant.  
The baseline $Acc_{bal}$ is $\approx66\%$ and increases to $\approx72\%$ for a $\epsilon_{rob} = 0.06$, and $AUC$ follows a similar trend.  
Similar to \census, the metrics for the attribute \sex have the opposite trend to the metrics for the attribute \race.  
However, it should be noted that the adversarial training in \lfw was not as effective as it was for \census.  
At $\epsilon_{rob} = 0.06$, $Acc_{rob}$ for $\modeldef$ is only $\approx33\%$.  
Although $Acc_{rob}$ for $\modeldef$ is higher than $\modelstd$, it would be a stretch to call this a robust model. 

Based on these experiments, changes in $Acc_{bal}$ and $AUC$ for a robust model fall well within the 5 p.p. range, indicating that~\ref{advtr} does not have a significant impact on~\ref{attinf}.  
However, because $\modeldef$ is not strongly robust for \lfw and the results are inconsistent across attributes and datasets, we cannot conclusively determine whether~\ref{advtr} affects~\ref{attinf}.

\begin{table*}[htb]
    \setlength\tabcolsep{2.5pt}
    \centering
    \small
    \caption{Interaction between~\ref{advtr} Adversarial Training and~\ref{modelext} Unauthorized Model Ownership. $Fid_{cor}$ is correct fidelity between $\modeldef$ and the stolen model $\modelstol$. Adversarial training raises fidelity but lowers correct fidelity on harder datasets.}

    \label{tab:advtr_modelext}
    \begin{tabular}{ c c c c c c c }
        \bottomrule
        
        \toprule
        \multicolumn{7}{c}{\textbf{\census}}\\
            & $Acc_{te}$ ($\modeldef$) & $Acc_{rob}$ ($\modelstd$) & $Acc_{rob}$ ($\modeldef$) & $\modelstol$: $Acc_{te}$ & $\modelstol$: $Fid$ & $\modelstol$: $Fid_{cor}$ \\
        \midrule
        Baseline ($\modelstd$) & 81.95~$\pm$~0.62 & - & - & - & - & - \\
        $\epsilon_{rob}$ = 0.01 & 82.27~$\pm$~1.59 & 73.25~$\pm$~1.15 & 81.43~$\pm$~1.84 & 82.74~$\pm$~1.69 & 95.30~$\pm$~0.98 & 84.10~$\pm$~1.47 \\
        $\epsilon_{rob}$ = 0.03 & 82.88~$\pm$~0.74 & 42.40~$\pm$~5.58 & 80.33~$\pm$~1.69 & 83.30~$\pm$~0.62 & 96.44~$\pm$~0.74 & 84.31~$\pm$~0.51 \\
        $\epsilon_{rob}$ = 0.05 & 83.02~$\pm$~0.89 & 26.02~$\pm$~4.46 & 78.99~$\pm$~2.91 & 83.46~$\pm$~0.88 & 96.95~$\pm$~0.60 & 84.28~$\pm$~0.76 \\
        $\epsilon_{rob}$ = 0.10 & 84.06~$\pm$~0.15 & 20.32~$\pm$~1.65 & 79.23~$\pm$~2.45 & 84.38~$\pm$~0.15 & 98.34~$\pm$~0.38 & 84.79~$\pm$~0.21 \\
        \midrule
        \multicolumn{7}{c}{\textbf{\fmnist}}\\
            & $Acc_{te}$ ($\modeldef$) & $Acc_{rob}$ ($\modelstd$) & $Acc_{rob}$ ($\modeldef$) & $\modelstol$: $Acc_{te}$ & $\modelstol$: $Fid$ & $\modelstol$: $Fid_{cor}$ \\
        \midrule
        Baseline ($\modelstd$) & 85.67~$\pm$~0.31 & - & - & - & - & - \\
        $\epsilon_{rob}$ = 0.01 & 84.81~$\pm$~0.69 & 77.38~$\pm$~0.73 & 80.63~$\pm$~0.69 & 84.45~$\pm$~0.60 & 92.42~$\pm$~0.45 & 88.27~$\pm$~0.59 \\
        $\epsilon_{rob}$ = 0.03 & 84.92~$\pm$~0.43 & 42.54~$\pm$~1.31 & 74.59~$\pm$~0.72 & 84.32~$\pm$~0.37 & 94.76~$\pm$~0.34 & 87.07~$\pm$~0.36 \\
        $\epsilon_{rob}$ = 0.05 & 83.86~$\pm$~0.58 & 19.59~$\pm$~0.91 & 69.08~$\pm$~1.46 & 83.26~$\pm$~0.43 & 95.78~$\pm$~0.23 & 85.39~$\pm$~0.38 \\
        $\epsilon_{rob}$ = 0.10 & 80.58~$\pm$~0.63 & 7.97~$\pm$~0.33 & 58.55~$\pm$~2.10 & 80.30~$\pm$~0.51 & 96.16~$\pm$~0.37 & 82.15~$\pm$~0.46 \\
        \midrule
        \multicolumn{7}{c}{\textbf{\lfw}}\\
            & $Acc_{te}$ ($\modeldef$) & $Acc_{rob}$ ($\modelstd$) & $Acc_{rob}$ ($\modeldef$) & $\modelstol$: $Acc_{te}$ & $\modelstol$: $Fid$ & $\modelstol$: $Fid_{cor}$ \\
        \midrule
        Baseline ($\modelstd$) & 82.89~$\pm$~0.63 & - & - & - & - & - \\
        $\epsilon_{rob}$ = 0.01 & 81.04~$\pm$~1.22 & 52.51~$\pm$~1.79 & 74.80~$\pm$~2.48 & 78.73~$\pm$~3.48 & 87.91~$\pm$~4.13 & 84.00~$\pm$~1.34 \\
        $\epsilon_{rob}$ = 0.03 & 77.83~$\pm$~1.62 & 18.03~$\pm$~0.65 & 55.08~$\pm$~9.00 & 77.84~$\pm$~1.91 & 89.11~$\pm$~3.48 & 81.31~$\pm$~2.61 \\
        $\epsilon_{rob}$ = 0.05 & 76.62~$\pm$~2.22 & 17.15~$\pm$~0.63 & 45.78~$\pm$~12.10 & 76.83~$\pm$~2.73 & 90.73~$\pm$~2.79 & 79.54~$\pm$~3.36 \\
        $\epsilon_{rob}$ = 0.10 & 65.99~$\pm$~0.61 & 17.12~$\pm$~0.63 & 59.68~$\pm$~13.60 & 65.97~$\pm$~0.67 & 99.74~$\pm$~0.30 & 66.02~$\pm$~0.68 \\
        \midrule
        \multicolumn{7}{c}{\textbf{\cifar}}\\
            & $Acc_{te}$ ($\modeldef$) & $Acc_{rob}$ ($\modelstd$) & $Acc_{rob}$ ($\modeldef$) & $\modelstol$: $Acc_{te}$ & $\modelstol$: $Fid$ & $\modelstol$: $Fid_{cor}$ \\
        \midrule
        Baseline ($\modelstd$) & 83.30~$\pm$~1.93 & - & - & - & - & - \\
        $\epsilon_{rob}$ = 0.01 & 80.85~$\pm$~0.26 & 23.82~$\pm$~1.82 & 66.82~$\pm$~0.57 & 79.30~$\pm$~0.77 & 83.94~$\pm$~0.85 & 88.14~$\pm$~0.42 \\
        $\epsilon_{rob}$ = 0.03 & 70.69~$\pm$~0.44 & 9.12~$\pm$~0.76 & 38.85~$\pm$~0.84 & 70.31~$\pm$~1.21 & 79.17~$\pm$~0.73 & 79.82~$\pm$~0.78 \\
        $\epsilon_{rob}$ = 0.05 & 66.31~$\pm$~0.78 & 7.96~$\pm$~1.23 & 26.86~$\pm$~0.58 & 66.18~$\pm$~0.80 & 78.94~$\pm$~1.09 & 74.78~$\pm$~0.76 \\
        $\epsilon_{rob}$ = 0.10 & 62.65~$\pm$~15.41 & 5.20~$\pm$~1.66 & 19.29~$\pm$~7.96 & 61.43~$\pm$~15.63 & 86.77~$\pm$~1.26 & 66.62~$\pm$~17.33 \\
        \bottomrule
        
        \toprule
    \end{tabular}
\end{table*}

\noindent\textbf{\ref{advtr} (Adversarial Training) and~\ref{modelext} (Unauthorized Model Ownership).}  
We evaluate how adversarial training (\ref{advtr}) influences unauthorized model ownership (\ref{modelext}), summarized in Table~\ref{tab:advtr_modelext}.  
In this setting, the adversary extracts $\modelstol$ from a robustly trained model $\modeldef$.  
We report results across four datasets and observe that the interaction varies by dataset.  

On \census, adversarial training has little effect on unauthorized model ownership.  
Across $\epsilon_{rob}=0.01$--$0.10$, $\modelstol$ fidelity increases slightly from $\approx95\%$ to $\approx98\%$, while correct fidelity remains stable, showing that the stolen model continues to mirror both the predictions and correctness of $\modeldef$.

For \fmnist, the trend is similar but slightly more pronounced.  
Fidelity rises to $\approx96\%$, while correct fidelity declines from $\approx88\%$ to $\approx82\%$ as robustness increases.  
Because $\modeldef$ continues to retain high test accuracy ($Acc_{te}$), $\modelstol$ remains a strong approximation, though its agreement on correct predictions weakens.  

On more complex datasets, the pattern shifts.  
For \lfw, $Fid$ rises from 88\% to nearly 100\%, but $Fid_{cor}$ drops from 84\% to 66\% as $\modeldef$ loses accuracy.  
On \cifar, the effect is clearer: at $\epsilon_{rob}=0.10$, $\modeldef$ has $Acc_{te}\approx62\%$ and $Acc_{rob}\approx19\%$, while $\modelstol$ attains similar $Acc_{te}$ ($\approx61\%$) and $Fid$ ($\approx87\%$), roughly $8$ p.p. more than when $\epsilon_{rob}=0.03$.  
Thus, the stolen model faithfully reproduces a boundary that is increasingly inaccurate.  

Overall, adversarial training often improves raw fidelity but reduces correct fidelity, especially on harder datasets.
This suggests it may \emph{increase} susceptibility to unauthorized model ownership by producing more predictable decision boundaries.
Further study with stronger defenses and modern architectures is needed to determine whether this effect is fundamental or an artifact of weaker training.

\begin{table*}[htbp]
\small
\caption{Interaction between~\ref{outrem} Outlier Removal and~\ref{modelext} Unauthorized Model Ownership. Removing outliers lowers $\modeldef$ accuracy but leaves stolen-model fidelity essentially unchanged.}
\label{tab:outrem_modext}
\begin{center}
\setlength\tabcolsep{2pt}
\begin{tabular}{ l c c c c c }
    \bottomrule

    \toprule
    & \multirow{2}{*}{$\modelstd$} & \multicolumn{4}{c}{$\modeldef$ (\% of outliers removed followed by retraining)} \\
        & & \textbf{10\%} & \textbf{20\%} & \textbf{30\%} & \textbf{40\%} \\
    \midrule
    \multicolumn{6}{c}{\textbf{\census}} \\
    $Acc_{te}$ & 84.60~$\pm$~0.16 & 81.80~$\pm$~0.30 & 80.73~$\pm$~0.52 & 79.43~$\pm$~0.45 & 77.61~$\pm$~0.39\\ 
    $\modelstol$: $Acc_{te}$ & 84.56~$\pm$~0.14 & 81.71~$\pm$~0.37 & 80.63~$\pm$~0.61 & 79.41~$\pm$~0.43 & 77.55~$\pm$~0.47\\  
    $\modelstol$: $Fid$ & 99.12~$\pm$~0.08 & 98.88~$\pm$~0.24 & 99.07~$\pm$~0.14 & 99.17~$\pm$~0.19 & 99.00~$\pm$~0.19\\ 
    $\modelstol$: $Fid_{cor}$ & 84.14~$\pm$~0.17 & 81.19~$\pm$~0.36 & 80.22~$\pm$~0.62 & 79.00~$\pm$~0.43 & 77.08~$\pm$~0.37\\ 
    \midrule

    \multicolumn{6}{c}{\textbf{\lfw}}\\
    $Acc_{te}$ & 81.40~$\pm$~0.83  & 80.78~$\pm$~1.19 & 79.41~$\pm$~1.65 & 77.61~$\pm$~2.78 & 78.25~$\pm$~0.29\\ 
    $\modelstol$: $Acc_{te}$ & 80.96~$\pm$~0.48 & 80.08~$\pm$~1.16 & 78.77~$\pm$~2.24 & 78.03~$\pm$~2.53 & 76.09~$\pm$~1.04\\  
    $\modelstol$: $Fid$ & 94.55~$\pm$~0.67 & 94.31~$\pm$~1.53 & 93.35~$\pm$~1.56 & 95.47~$\pm$~0.59 & 93.46~$\pm$~2.37 \\ 
    $\modelstol$: $Fid_{cor}$ & 78.46~$\pm$~0.38 & 77.58~$\pm$~0.66 & 75.76~$\pm$~1.96 & 75.56~$\pm$~2.56 & 73.90~$\pm$~1.64\\ 
    \midrule

    \multicolumn{6}{c}{\textbf{\fmnist}}\\
    $Acc_{te}$ & 83.51~$\pm$~0.55 & 79.19~$\pm$~0.58 & 78.40~$\pm$~1.09 & 75.39~$\pm$~1.22 & 75.22~$\pm$~1.27\\ 
    $\modelstol$: $Acc_{te}$ & 82.74~$\pm$~0.66 & 79.61~$\pm$~0.67 & 79.28~$\pm$~0.45 & 74.79~$\pm$~1.46 & 75.65~$\pm$~1.58\\  
    $\modelstol$: $Fid$ & 92.89~$\pm$~0.45  & 90.61~$\pm$~0.34 & 89.68~$\pm$~0.56 & 89.72~$\pm$~0.59 & 88.75~$\pm$~0.85\\ 
    $\modelstol$: $Fid_{cor}$ & 80.27~$\pm$~0.58 & 75.65~$\pm$~0.70 & 74.66~$\pm$~0.90 & 71.03~$\pm$~1.48 & 70.91~$\pm$~1.62\\  
    \midrule

    \multicolumn{6}{c}{\textbf{\cifar}}\\
    $Acc_{te}$ &  83.65~$\pm$~0.38 & 75.35~$\pm$~0.33 & 73.86~$\pm$~1.32 & 71.86~$\pm$~0.67 & 70.50~$\pm$~1.01\\ 
    $\modelstol$: $Acc_{te}$ & 82.03~$\pm$~0.56 & 75.81~$\pm$~0.48 & 74.26~$\pm$~1.32 & 72.13~$\pm$~1.11 & 71.62~$\pm$~1.25\\  
    $\modelstol$: $Fid$ & 86.52~$\pm$~0.32 & 84.66~$\pm$~0.76 & 84.14~$\pm$~1.37 & 83.83~$\pm$~0.49 & 83.61~$\pm$~0.89\\ 
    $\modelstol$: $Fid_{cor}$ & 77.73~$\pm$~0.54 & 70.17~$\pm$~0.41 & 68.54~$\pm$~1.68 & 66.49~$\pm$~0.92 & 65.43~$\pm$~1.26\\ 
    \bottomrule

    \toprule
\end{tabular}
\end{center}
\end{table*}

\noindent\textbf{\ref{outrem} (Outlier Removal) and~\ref{modelext} (Unauthorized Model Ownership).}
Finally, we evaluate the interaction between outlier removal (\ref{outrem}) and unauthorized model ownership (\ref{modelext}).

\begin{figure}[t]
    \centering
    \includegraphics[width=\linewidth]{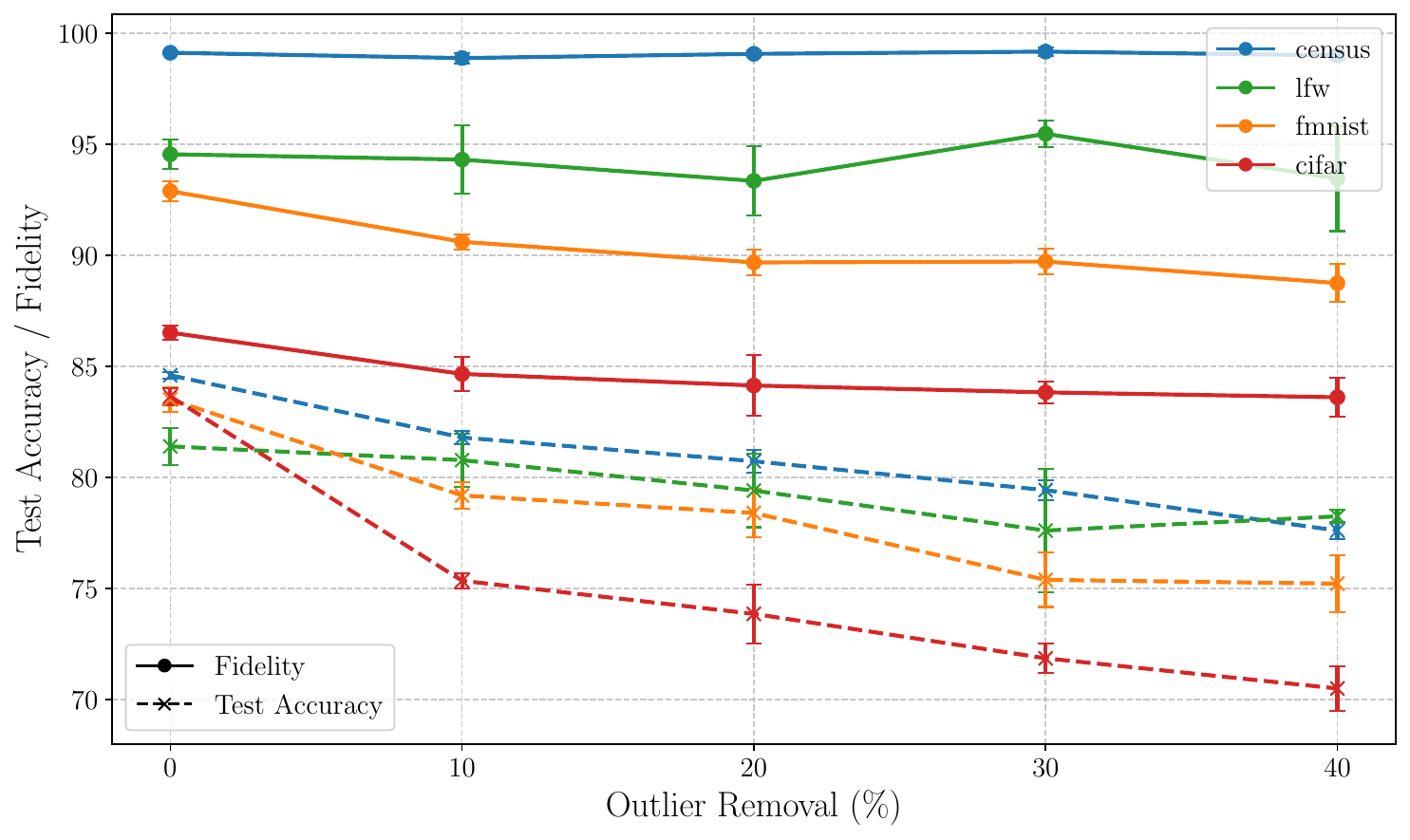}
    \caption{Comparison of $Acc_{te}$ of $\modeldef$ and $Fid$ between $\modeldef$ and $\modelstol$ for the given percentage of outliers removed from $\modeldef$.
    $Acc_{te}$ shows a clear downward trend, while fidelity is not affected.
    The colors differentiate the dataset, while the dotted lines represent $Acc_{te}$ and the solid lines represent $Fid$. }
    \label{fig:outrem_fid}
    \Description{Line plot of defended-model test accuracy and stolen-model fidelity against the percentage of outliers removed, for three datasets. Test accuracy trends downward as more outliers are removed, while fidelity stays roughly constant.}
\end{figure}

Outlier removal trains the model after discarding training records with the lowest influence, estimated using $k$-nearest-neighbor Shapley values~\cite{jia2019EfficientTaskspecificData}.
These scores quantify how much each training record contributes to model predictions on a held-out set, allowing us to iteratively remove the least influential records before retraining the model on the remaining data.

Our results show that as the percentage of outliers removed increases, the test accuracy of $\modeldef$ decreases.
This is expected since~\ref{outrem} uses Shapley values to identify data records with the highest influence, and removing them can degrade accuracy~\cite{jia2019EfficientTaskspecificData}.
For example, on \census, $Acc_{te}$ drops from $\approx84.6\%$ (baseline) to $\approx81\%$ with $10\%$ outliers removed, and further to $\approx77\%$ at $40\%$.

\begin{figure}[ht]
    \centering
    \includegraphics[width=\linewidth]{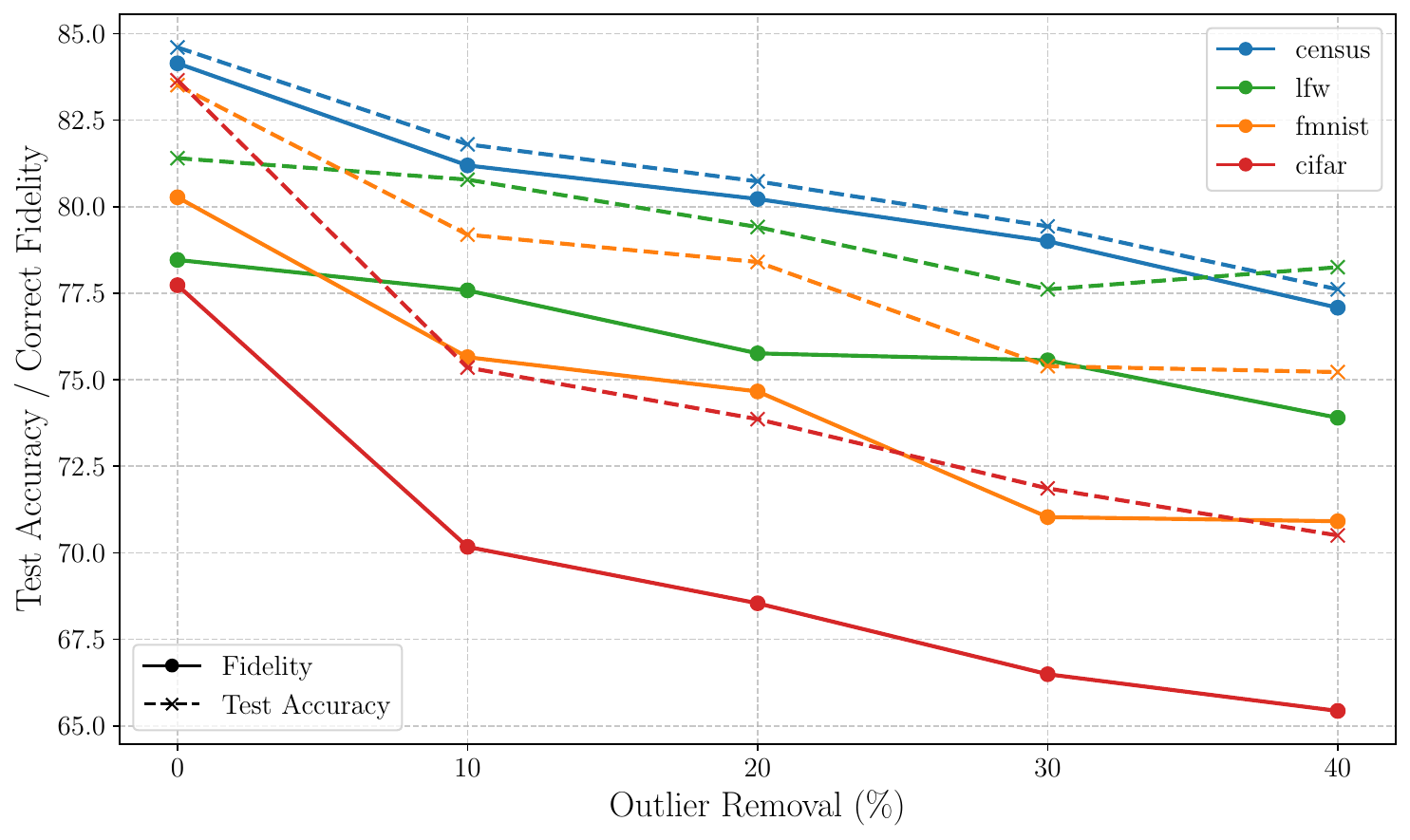}
    \caption{Comparison of $Acc_{te}$ of $\modeldef$ and $Fid_{cor}$ between $\modeldef$ and $\modelstol$ for the given percentage of outliers removed from $\modeldef$.
    Both measures follow a similar downward trend.
    Solid lines represent $Fid_{cor}$. Error bars omitted for readability.}
    \label{fig:outrem_cor_fid}
    \Description{Line plot of defended-model test accuracy and correct-label fidelity against the percentage of outliers removed, for three datasets. Both measures follow a similar downward trend as more outliers are removed.}
\end{figure}

Analyzing the effect this has on unauthorized model ownership, Figure~\ref{fig:outrem_fid} shows that while $Acc_{te}$ of $\modeldef$ drops significantly for all datasets, the fidelity of $\modelstol$ stays constant (in the case of \census) or drops slightly (in the case of \cifar and \fmnist).
This suggests that outlier removal does not affect unauthorized model ownership.

Figure~\ref{fig:outrem_cor_fid} compares $Fid_{cor}$ with $Acc_{te}$, revealing that both trend downward.
For \cifar, removing $10\%$ of the outliers causes a $\geq7.5$~p.p. drop in both $Acc_{te}$ and $Fid_{cor}$.
This indicates that while overall fidelity is preserved, correct fidelity is sensitive to the degradation of $\modeldef${'s} accuracy.
We conclude that~\ref{outrem} does not meaningfully affect~\ref{modelext}.

\begin{takeaway}
The strength and direction of unintended interactions vary greatly with the specific dataset, model, and hyperparameters, making direct measurement essential.
\end{takeaway}

\section{Discussion}\label{sec:discussion}
\noindent\textbf{Adding more attacks and defenses.}
While \method covers a wide range of risks, we prioritized algorithms with open-source implementations that have been analyzed in multiple peer-reviewed articles -- each requiring rigorous testing and validation across datasets and architectures to ensure consistent, reproducible behavior.  
Community contributions of state-of-the-art attacks and defenses would further extend this foundation, reducing the significant engineering effort currently required to adapt and validate each module and making \method even more useful for studying unintended interactions.

\noindent\textbf{Studying conflicting defenses.}
While \method is well suited for evaluating how a single defense affects an unrelated risk, it remains difficult to study how multiple defenses interact when deployed together.  
For example, both adversarial training and DP-SGD are applied during training, which is not yet supported by \method.  
This limitation is important for real-world deployments, where practitioners often need to defend models against several risks simultaneously~\cite{szyller2023ConflictingInteractionsProtection}.  
We leave this to future work, as \method remains under active development.

Similarly, the effect of adversarial training~(\ref{advtr}) on fingerprinting~(\ref{fngrprnt}) warrants further study.  
Fingerprinting techniques that exploit decision boundaries (e.g.,~\cite{maini2021DatasetInferenceOwnership}) have been shown to fail in the presence of~\ref{advtr}~\cite{szyller2023RobustnessDatasetInference}.  
Other techniques that rely on conferrable adversarial examples (e.g.,~\cite{lukas2021DeepNeuralNetwork}) would also be affected by~\ref{advtr}~\cite{szyller2023ConflictingInteractionsProtection}.

\noindent\textbf{Summary.}
This work introduces \method, a Python library designed to systematically evaluate both \textit{intended} and \textit{unintended} interactions among machine learning defenses and risks. 
We first identified the design requirements for such a system, including comprehensiveness~(\ref{comprehensive}), consistency~(\ref{consistent}), extensibility~(\ref{extensible}), and applicability~(\ref{applicable}), and then implemented \method to meet these requirements through a modular, risk-centric architecture. 
Using \method, we demonstrated several previously unexplored interactions as proof of concept, showing how \method serves as a foundation for future studies.

\method is a step toward unifying fragmented efforts in evaluating trustworthy machine learning.  
As machine learning continues to expand to large-scale foundation and language models, understanding how defenses interact across multiple risks is increasingly critical.  
\method covers canonical vision and tabular models to ensure comparability with prior work, and Section~\ref{sec:extensibility} shows the same evaluations transferring to text with a three-billion-parameter LLM victim.
We hope that through open source, \method can grow to serve all researchers and practitioners in this space.

\bibliographystyle{ACM-Reference-Format}
\bibliography{paper}

\appendix
\section{Metric Definitions}
\label{sec:metric-definitions}

Several risks in \method are evaluated using the same underlying metric, applied to different data distributions (e.g., clean, adversarial, or poisoned inputs).
To avoid redundancy, we define each metric once and specify how it is instantiated for different risks.

\subsection{Classification Metrics}

Classification accuracy is defined as
\begin{equation}
\label{eq:acc}
\mathrm{Acc} = \frac{1}{|\mathcal{D}|} \sum_{(x,y)\in\mathcal{D}} \mathbb{I}\!\left[\arg\max f(x) = y\right],
\end{equation}
where $f$ denotes the model and $\mathcal{D}$ the evaluation dataset.
Robust accuracy evaluates Eq.~\ref{eq:acc} on adversarially perturbed inputs, poisoned data accuracy evaluates it on poisoned records, and attack accuracy evaluates it on records generated by the corresponding attack.
In all experiments, accuracy is reported as a percentage.

Balanced accuracy is defined as
\begin{equation}
\label{eq:balacc}
Acc_{bal} = \frac{1}{2}\left(
\frac{\mathrm{TP}}{\mathrm{TP}+\mathrm{FN}} +
\frac{\mathrm{TN}}{\mathrm{TN}+\mathrm{FP}}
\right),
\end{equation}
and we report the maximum balanced accuracy over all decision thresholds.
This metric is used for membership and attribute inference to account for class imbalance.

The area under the receiver operating characteristic curve (ROC AUC) is defined as
\begin{equation}
\label{eq:auc}
\mathrm{AUC} = \int_{0}^{1} \mathrm{TPR}(\mathrm{FPR}) \, d(\mathrm{FPR}),
\end{equation}
measuring the trade-off between true positive and false positive rates across decision thresholds.

TPR@1\%FPR is defined as $\mathrm{TPR}|_{\mathrm{FPR}\leq 0.01}$, the true positive rate evaluated at the largest false positive rate not exceeding 1\%.

\subsection{Reconstruction Metrics}

For data reconstruction, we compute the mean squared error between the reconstructed record for each class and the average data record of that class, and report the average MSE across classes:
\begin{equation}
\label{eq:mse}
\mathrm{MSE} = \frac{1}{C} \sum_{c=1}^{C} \lVert \hat{x}_c - \bar{x}_c \rVert_2^2,
\end{equation}
where $\hat{x}_c$ denotes the reconstructed record for class $c$ and $\bar{x}_c$ the average data record of that class.

Structural similarity index measure (SSIM) is defined as~\cite{wang2004ImageQualityAssessment}
\begin{equation}
\label{eq:ssim}
\mathrm{SSIM}(x,\hat{x}) =
\frac{(2\mu_x\mu_{\hat{x}}+c_1)(2\sigma_{x\hat{x}}+c_2)}
{(\mu_x^2+\mu_{\hat{x}}^2+c_1)(\sigma_x^2+\sigma_{\hat{x}}^2+c_2)},
\end{equation}
where $\mu$, $\sigma$, and $\sigma_{x\hat{x}}$ denote sample means, variances, and covariance, respectively.
We compute SSIM between reconstructed records and average data records, and report the mean SSIM across classes.

\subsection{Fairness Metrics}

Fairness metrics compare model behavior across protected subgroups $a \in \mathcal{A}$.

Demographic parity is measured as the absolute difference $\mathrm{DP} = \left| P(f(x)=1 \mid a=0) - P(f(x)=1 \mid a=1) \right|$.

True positive parity and false positive parity are defined as
\begin{equation}
\label{eq:tpp}
\begin{aligned}
\mathrm{TPP} =
\Big|
& P(f(x)=1 \mid y=1, a=0) \\
& -\; P(f(x)=1 \mid y=1, a=1)
\Big|,
\end{aligned}
\end{equation}
\begin{equation}
\label{eq:fpp}
\begin{aligned}
\mathrm{FPP} =
\Big|
& P(f(x)=1 \mid y=0, a=0) \\
& -\; P(f(x)=1 \mid y=0, a=1)
\Big|,
\end{aligned}
\end{equation}
respectively.

Equalized odds is computed as the sum of true positive parity and false positive parity, $\mathrm{EO} = \mathrm{TPP} + \mathrm{FPP}$.

The $p$-rule is defined as
\begin{equation}
\label{eq:prule}
\min\left(
\frac{P(f(x)=1 \mid a=1)}{P(f(x)=1 \mid a=0)},
\frac{P(f(x)=1 \mid a=0)}{P(f(x)=1 \mid a=1)}
\right),
\end{equation}
and is reported as a percentage.

\section{Use of Generative AI}
AI tools were used in the development of \method and as a writing assistant for this paper. Specifically, \method has been in development since 2019. As such, the initial library was written entirely by hand. The most recent development used AI tools to help. The initial few drafts of this manuscript were entirely written by hand. AI was only used for feedback and to refine specific sentences.

\end{document}